# Ge epitaxy at ultra-low growth temperatures enabled by a pristine growth environment

*Christoph Wilflingseder, Johannes Aberl, Enrique Prado Navarette, Günter Hesser, Heiko Groiss, Maciej O. Liedke, Maik Butterling, Andreas Wagner, Eric Hirschmann, Cedric Corley-Wiciak, Marvin H. Zoellner, Giovanni Capellini, Thomas Fromherz, Moritz Brehm*

Christoph Wilflingseder, Johannes Aberl, Enrique P. Navarette, Thomas Fromherz, Moritz Brehm - Institute of Semiconductor and Solid State Physics, Johannes Kepler University Linz, Altenberger Straße 69, 4040, Linz, Austria

Günter Hesser, Heiko Groiss - Christian Doppler Laboratory for Nanoscale Phase Transformations, Center for Surface And Nanoanalytics (ZONA), Johannes Kepler University Linz, Altenberger Straße 69, 4040, Linz, Austria

Maciej O. Liedke, Maik Butterling, Andreas Wagner, Eric Hirschmann - Helmholtz-Zentrum Dresden-Rossendorf e.V., Institute of Radiation Physics, Dresden, 01328, Germany

Cedric Corley-Wiciak - ESRF – European Synchrotron Radiation Facility, 71, Avenue des Martyrs, CS 40220, 38043 Grenoble Cedex 9, France

Marvin H. Zoellner, Giovanni Capellini - IHP – Leibniz-Institut für innovative Mikroelektronik, Im Technologiepark 25, D-15236, Frankfurt(Oder), Germany

Giovanni Capellini - Dipartimento di Scienze, Università Roma Tre, V.le G. Marconi 446, 00146 Roma, Italy

KEYWORDS. germanium, silicon, epitaxy, strain, transmission electron microscopy, nanobeam x-ray diffraction, positron annihilation lifetime spectroscopy

## Abstract

**Germanium (Ge), the next-in-line group-IV material, bears great potential to add functionality and performance to next-generation nanoelectronics and solid-state quantum**

transport based on silicon (Si) technology. Here, we investigate the direct epitaxial growth of two-dimensional high-quality crystalline Ge layers on Si deposited at ultra-low growth temperatures ($T_{Ge}$ = 100°C - 350°C) and pristine growth pressures (≲$10^{-10}$ mbar). First, we show that $T_{Ge}$ does not degrade the crystal quality of homoepitaxial Ge/Ge(001) by comparing the point defect density using positron annihilation lifetime spectroscopy. Subsequently, we present a systematic investigation of the Ge/Si(001) heteroepitaxy, varying the Ge coverage ($\Theta_{Ge}$, 1, 2, 4, 8, 12, and 16 nm) and $T_{Ge}$ (100°C to 300°C, in increments of 50°C) to assess the influence of these parameters on the layer's structural quality. Atomic force microscopy revealed a rippled surface topography with superimposed grainy features and the absence of three-dimensional structures, such as quantum dots. Transmission electron microscopy unveiled pseudomorphic, grains of highly crystalline growth separated by defective domains. Thanks to nanobeam scanning x-ray diffraction measurements, we were able to evidence the lattice strain fluctuations due to the ripple-like structure of the layers. We conclude that the heteroepitaxial strain contributes to the formation of the ripples, which originate from the kinetic limitations of the ultra-low temperatures.

## 1. Introduction

Silicon is the most widely used material for electronic devices, but for further technological progress, devices must get smaller, faster, more versatile, or less power-consuming. Since scaling down Si-based devices is becoming increasingly challenging, carefully chosen additions of materials have been explored. Ge stands out because of its high charge carrier mobility and lower band gap. Seminal work has been performed to integrate Ge on Si [1-5], for decades. Most implementations focused on Ge on Si and silicon-on-insulator (SOI) [6,7] based on thick, strain-relaxed SiGe virtual substrates [8-11]. Using such strain engineering techniques and Ge as a channel material, it was shown to be possible to enhance the hole mobility to a record value of ~4·$10^6$ $\frac{cm^2}{Vs}$ [12]. Ge, in general, broadens the spectrum of Si-based applications. With a bandgap within the telecommunications C-band, Ge is the base for new perspectives in Si photonics, such as lasers, modulators, waveguides, and spintronics [13-19]. On another hand, compressively strained Ge layers emerged as a promising platform for solid-state quantum computing, and their co-integration on Si substrates offers a pathway toward a scaled, functional quantum processor [20]. While all these devices rely on thick, relaxed, and thus defective Ge and SiGe buffer layers, recently, an alternative approach was demonstrated. Strained and two-dimensional SiGe and Ge nanosheets were directly grown on Si and SOI substrates using low temperature epitaxy [21,22]. This method enabled the top-down fabrication of versatile, high-quality nanoelectronics devices [23] such as reconfigurable transistors and devices based on negative differential resistance [24-27]. Besides these devices' excellent electronic characteristics [24-27], detailed investigations of the critical growth parameters such as temperature, layer thickness, and limits of the strained growth have not been addressed. Due to the ~4.2% larger lattice constant of Ge compared to Si, only a ~0.5 nm thick Ge-wetting layer (WL) can be directly grown on a Si substrate for typical epitaxy temperature (500°C < $T_{Ge}$ < 800°C), [28,29]. Beyond that thickness, the Stranski-Krastanow growth dynamics impose the formation of elastically relaxed 3D islands (usually referred to as quantum-dot, QD) that eventually undergo a plastic relaxation process [30,31]. Lowering $T_{Ge}$ limits QD formation kinetically, as demonstrated by Bean *et al*. [32] about 40 years ago by separating the growth of Si buffer and Ge epilayer into two steps with different growth temperatures and reducing the $T_{Ge}$ of the epilayer to 400°C. In 1991, Eaglesham and Cerullo [33] grew planar Ge layers on Si at $T_{Ge}$s as low as 50°C, discovering that films thicker than 3.5 nm were partially plastically relaxed, thanks to the formation of misfit dislocations MDs. Within the $T_{Ge}$ range of 50-150°C, there is a maximum epilayer thickness that can be deposited, beyond which the growth resulted in the formation of an amorphous layer. Indeed, at lower temperatures the ad-atom surface diffusion on Si and on the Ge WL is reduced and residual atoms and molecules originating from the chamber background and the sources can be incorporated during growth [34], leading to poor epitaxy.

In this work, we show the often-neglected role of the deep-ultra-high vacuum conditions ($\leq 2.0 \cdot 10^{-10}$ mbar) in a molecular-beam epitaxy (MBE) chamber as a key enabler for expanding the growth parameters space of strained, epitaxial Ge thin layers deposited either on Ge or Si substrates. We first demonstrate that the crystalline quality of Ge/Ge(001) epilayers is almost unaffected by the deposition temperature $T_{Ge}$ in the range 100°C – 350°C as evidenced by variable energy positron annihilation lifetime spectroscopy measurements (VEPALS). Subsequently, we have further rigorously studied the strained layer growth of Ge on Si, as a function of $T_{Ge}$ and Ge layer thickness $\Theta_{Ge}$. Depending on $T_{Ge}$, the formation of large QDs in strained heteroepitaxy is suppressed, and the layers build domains of highly crystalline regions, separated by thin areas of distorted growth. The findings are confirmed through various analytical techniques, including variable energy positron annihilation lifetime spectroscopy (VEPALS), atomic force microscopy (AFM), transmission electron microscopy (TEM), x-ray diffraction (XRD), and scanning x-ray diffraction microscopy (SXDM). Consequently, ultra-low temperature (ULT ≡ <350°C) Ge epitaxy on Si(001) should be considered for further research with the aim of implementing it in industry MBE systems for next-generation nanoelectronics.

## 2. Experimental Methods
### 2.1. Ultra-low temperature epitaxy
#### 2.1.1. Ge on Ge(001)

All samples were grown in a Riber SIVA-45 solid source MBE system. For Ge homoepitaxy, Cz Ge(001) substrates were cleaned using a plasma cleaner and a UV ozone cleaner, with intermediate immersion in solvents within an ultrasonic bath. Subsequently, the substrates were degassed at 300°C for 30 min and the oxide was removed through thermal desorption (750°C for 10 min). Hereafter, for all samples, a 50 nm thick Ge buffer was grown at a $T_{Ge}$ = 320°C and at a rate of 0.02 nm/s. Within growth interrupts of 7 min, 16 min, 40 min, and 53 min, the sample temperature was ramped to 350°C, 200°C, 150°C and 100°C, at which 50 nm of Ge were deposited at a growth rate of 0.02 nm/s.

#### 2.1.2. Ge on Si(001)

The heteroepitaxial and strained $\Theta_{Ge}$ and $T_{Ge}$ series were grown on high-resistivity, 4-inch, and intrinsic FZ Si(001) substrates. After wafer cleaning, including an RCA (Radio Corporation of America) cleaning process, the substrates were submerged for 1 min in diluted hydrofluoric acid (HF 1%) to remove the native oxide. Subsequently, the substrates underwent a two-step degassing process: 15 min at 700°C and 30 min at 450°C, before we grew a 75.5 nm thick Si buffer at growth temperatures that were linearly decreased from 650°C to 600°C, while the deposition rate was increased from 0.05 nm/s to 0.075 nm/s. Next, the substrates were cooled down to $T_{Ge}$, i.e., 100°C, 150°C, 200°C, 250°C, and 300°C, respectively. Depending on $T_{Ge}$, the growth interruption varied from 65 min (for 100°C) to 13 min (300°C). To increase the size of our sample set in controlled conditions, we created on each substrate three areas receiving different $\Theta_{Ge}$ by switching off the substrate rotation and using a manual shutter to partially cover the substrate. In this way, a comprehensive matrix of 10 wafers was grown, featuring 6 varying Ge thicknesses $\Theta_{Ge}$ of 1 nm, 2 nm, and 4 nm, as well as 8 nm, 12 nm, and 16 nm for each of the five $T_{Ge}$s. All samples remained uncapped. Two selected growth log files can be found in the supplementary material (see Figure S. 2(a) and (b)), showcasing the highest and lowest growth pressure ($p_{Ge}$) on average during Ge deposition (~2·10$^{-10}$ mbar, and ~9·10$^{-11}$ mbar, respectively) for all samples.

## 2.2. Layer characterization

### 2.2.1. Variable Energy Positron Annihilation Lifetime Spectroscopy

VEPALS measurements were conducted at the mono-energetic positron source (MePS) beamline at HZDR, Germany [35]. A CeBr$_3$ scintillator detector coupled to a Hamamatsu R13089-100 photomultiplier tube (PMT) was utilized for gamma photon detection. The signals were processed by the SPDevices ADQ14DC-2X digitizer (14 bit vertical resolution and 2GS/s horizontal resolution) [36]. The overall time resolution of the measurement system is approximately 0.25 ns and all spectra contained at least $1·10^7$ counts. A typical lifetime spectrum $N(t)$, the absolute value of the time derivative of the positron decay spectrum, is described by

$$N(t) = R(t) * \sum_{i=1}^{k+1} \frac{I_i}{\tau_i} e^{\frac{-t}{\tau_i}} + \text{Background} \tag{1}$$

where $k$ different defect types contributing to the positron trapping are related to $k + 1$ components in the spectra with the individual positron lifetimes $\tau_i$ and intensities $I_i$ ($\Sigma I_i=1$) [35]. The instrument resolution function $R(t)$ is a sum of two Gaussian functions with distinct intensities and relative shifts depending on the positron implantation energy, $E_p$. It was determined by the measurement and analysis of a reference sample, i.e. amorphous Yttria-Stabilized Zirconia (YSZ), which exhibited a single well-known lifetime component of ~0.182 ns. The background was negligible; hence, it was fixed to zero. All the spectra were deconvoluted using a non-linear least-squares fitting method, minimized by the Levenberg-Marquardt algorithm, employed within the fitting software package PALSfit [37] into two major lifetime components, which directly evidence localized annihilation at two different defect types (sizes; $\tau_1$ and $\tau_2$). Their relative intensities scale typically with the concentration of each defect type. In general, positron lifetime increases with defect size and open volume size. The positron lifetime and its intensity have been probed as a function of the positron implantation energy $E_p$ which was recalculated to the mean implantation depth $\langle z \rangle$. The average positron lifetime $\tau_{av}$ is defined as $\tau_{av} = \sum_i \tau_i \cdot I_i$.

### 2.2.2. Atomic Force Microscopy

The surface topography of the samples was studied using AFM, using a Veeco Dimension 3100 AFM, equipped with OMCL-AC160TS-R3 cantilevers (Olympus Corporation), which had a probe radius of 7 nm. Operating in tapping mode, 1×1 µm² (see Figure 2(a)) and 5×5 µm² (see Figure S. 3 micrographs with resolutions of 512 pixels/line were captured to determine both large-scale and small-scale features of the samples.

### 2.2.3. Transmission Electron Microscopy

The TEM experiments were carried out in a JEOL JEM-2200FS (JEOL, Japan) operated at an acceleration voltage of 200 kV. The TEM is equipped with an in-column Ω-filter and a TemCam-XF416 (TVIPS, Germany) CMOS-based camera. The conventional TEM investigations include high-resolution (HR)TEM as well as bright field (BF) and dark field (DF) imaging at two-beam conditions (TBC) of various diffraction reflexes. Plan-view (for examinations near the [001] zone axis), as well as cross-sectional (for examinations near a <110> zone axis) specimens were prepared classically by mechanical polishing (dimpling or wedge-polishing), followed by a final Ar-sputtering step to achieve electron transparency. Additionally, a ZEISS Crossbeam 1540XB (ZEISS, Germany) scanning electron microscope (SEM) with a focused ion beam (FIB) add-on was used to prepare the ready-to-use cross-sectional TEM lamellae.

### 2.2.4. X-ray Diffraction

Laboratory XRD measurements were performed with a *Rigaku Smartlab* diffractometer with a rotating-anode Cu Kα source, a Ge(400) × 2 channel-cut beam monochromator, and an x-ray area detector. For each sample, we recorded a Reciprocal Space Map (RSM) around the 004 and 224 Bragg reflection of the Ge QW layer to calculate the in-plane lattice strain $\varepsilon_{xx} = \varepsilon_{yy}$ and the out-of-plane strain $\varepsilon_{zz}$. These are the diagonal components of the strain tensor, which are linked through equation (2) by the Poisson number $v_{13}$ [38]:

$$\varepsilon_{zz} = -\frac{(\varepsilon_{xx} + \varepsilon_{yy})}{\frac{1}{v_{13}} - 1} \quad (2)$$

We note that this equation is valid only in the assumption of no surface normal stress ($\sigma_{zz} = 0$), however, this approximation holds well for thin epitaxial layers near a free surface. We also determine the degree of relaxation according to equation 3:

$$R_D = \frac{a_{\parallel} - a_{Si}}{a_{Ge} - a_{Si}} \qquad (3)$$

Here, $a_{\parallel}$ refers to the in-plane lattice constant of the Ge layer determined from XRD, while $a_{Si}$ and $a_{Ge}$ are the literature values for the lattice constants of Si and Ge, respectively. Laboratory XRD yields the strain state of the Ge QW layer averaged over a wide area on the sample due to the large width of the parallel x-ray beam. To probe the local strain landscape with fine spatial resolution, we employed an advanced synchrotron-based technique, scanning x-ray diffraction microscopy (SXDM) [39], at the hard x-ray nanoprobe beamline ID01/ESRF [40]. The energy was set at 9 keV, and the beam was focused by a Fresnel zone plate (FZP) to a focal point of 25 nm. Diffraction maps of the 004 Bragg reflection from the Ge QW layer were recorded by scanning the sample across the beam in a $(x, y)$ raster, while the intensity of the scattered x-rays was recorded continuously on a *Maxipix* area detector. To sample the 3D reciprocal space, these maps were measured for a series of rocking angles $\omega$, yielding a five-dimensional (5D) dataset. The diffraction data was analyzed with the *SXDM* and *Xrayutilities* packages for *Python*, providing finely resolved maps of the local scattering vector $Q_{004}$ for the 004 Bragg reflection, the $c$ lattice parameter and the vertical strain $\varepsilon_{zz}$ according to equations (4-6)

$$Q_{004} = \begin{bmatrix} Q_z \\ Q_y \\ Q_x \end{bmatrix} \qquad (4)$$

$$c = \frac{1}{|Q_{004}|} \qquad (5)$$

$$\varepsilon_{zz} = \frac{c}{a_0} - 1 \qquad (6)$$

Moreover, the lattice rotation $w_{yz}$ is calculated as the tilt of the scattering vector in reciprocal space according to equation 7:

$$w_{yz} = \tan^{-1}\left(\frac{Q_y}{Q_z}\right) \qquad (7)$$

## 2.3. Results and Discussion
### 2.3.1. Ge on Ge(001)

At first, we investigated the influence of the $T_{Ge}$ on homoepitaxial Ge/Ge(001), thereby excluding any effects attributable to heteroepitaxial strain. Therefore, the defect microstructure in the Ge layers was evaluated using positron annihilation lifetime spectroscopy (PALS). The decomposition of the experimental positron lifetime spectra revealed two major defect contributions, the positron lifetime components $\tau_1$ and $\tau_2$. The shortest lifetime $\tau_1$ represents positron trapping and annihilation with electrons inside of small vacancy like defects (single vacancies), whereas $\tau_2$ originates from defect states in the sub-surface region of the films (vacancy agglomerations). In the absence of a sufficient number of traps, positrons can diffuse freely reaching eventually the surface or bulk [41]. In Figure 1(a) we show the lifetimes $\tau_1$ and $\tau_2$ as a function of the $T_{Ge}$ for a series of nominally identically thick Ge layers. In addition, the average lifetime $\tau_{av}$ is plotted, which is a parameter sensitive to the overall defect size. For the $\tau_{av}$, the positron lifetime components are combined and weighted according to their relative intensities. These intensities are directly correlated with the concentration of defects and are presented in Figure 1(b). The PALS analysis shows a monotonic decrease of positron lifetime with the implantation depth, with a minimal influence of the deposition temperature. The decrease of positron lifetime across the depth is a consequence of a moderate number of positron traps (point defects, $\tau_1$), which could hinder positron back diffusion to the surface ($\tau_2$), as well as a positron implantation profile, which broadens with its kinetic energy, i.e. depth [42]. The increase of $\tau_1$ with deposition temperature indicates an onset of agglomeration of single point defects, preexisting in the 100°C sample, and likely originates from the increased point defect density ($I_1$ raises with increased $T_{Ge}$ in Figure 1(b)), hence a smaller fraction of positrons ($\tau_1$) can reach and annihilate with surface states ($\tau_2$). We can safely assume that a shorter lifetime $\tau_1$ represents a mixture of Ge monovacancy (V$_{Ge}$) [43] with a small fraction of bivacancies (2×V$_{Ge}$), and a longer $\tau_2$ arises from larger vacancy agglomerations (about four vacancies within a complex 4×V$_{Ge}$) in the sub-surface region (see Figure 1(a)). In general, there is virtually no difference in the point defect concentration depending on the growth temperature, since the relative intensity does not change for larger temperatures. We notice that both $\tau_1$ and $\tau_2$, although irregularly, tend to mildly increase with $T_{Ge}$, pointing to an increasing fraction of bivacancies and to larger vacancy agglomerations, respectively [43]. On the other hand, the initial concentration of $V_{Ge}$ ($I_1\approx65\%$) rises to $I_1\approx78\%$ with $T_{Ge}$ (see Figure 1(b)), which indicates larger trapping at small point defects and an increase of their density. At the same time, $I_2$ concomitantly decreases as a smaller fraction of positrons can arrive into the sub-surface region.

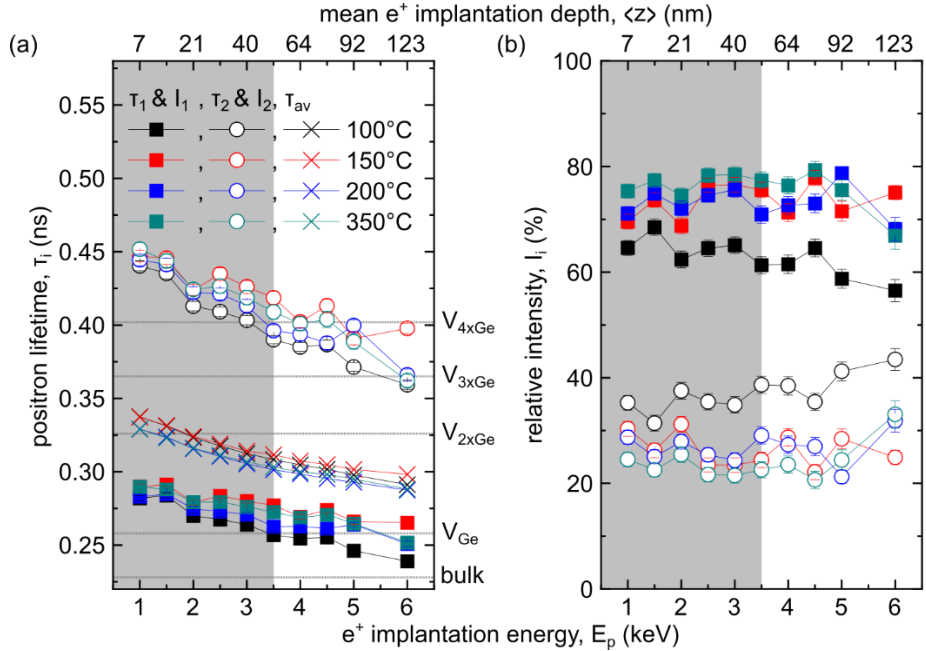

**Figure 1.** (a) Positron annihilation lifetime spectroscopy depth profiles of positron lifetimes $\tau_1$, $\tau_2$, $\tau_{av.}$ and (b) their relative intensities $I_1$, $I_2$ for Ge layers MBE deposited at 100°C, 150°C, 200°C, and 350°C. The horizontal dotted lines denote calculated defect states for a Ge crystal [43]. The grey areas denote the region of the top Ge layer.

In summary, the PALS results demonstrate that for the unstrained growth of Ge on Ge in deep UHV, the growth temperature can be lowered from 350°C to 100°C without an increase in the density of point defects. We note that also the significantly longer growth interrupt for lower $T_{Ge}$ had no detrimental side effects on the layer quality. This finding contradicts the conventional wisdom that even homoepitaxial growth at very low temperatures eventually breaks down to form a polycrystalline or an amorphous phase [33]. In turn, these results strongly indicate that a loss of crystallinity is instead a consequence of impurity incorporation due to poor $p_{Ge}$s and lack of contaminant desorption.

### 2.3.2. Ge on Si(001)

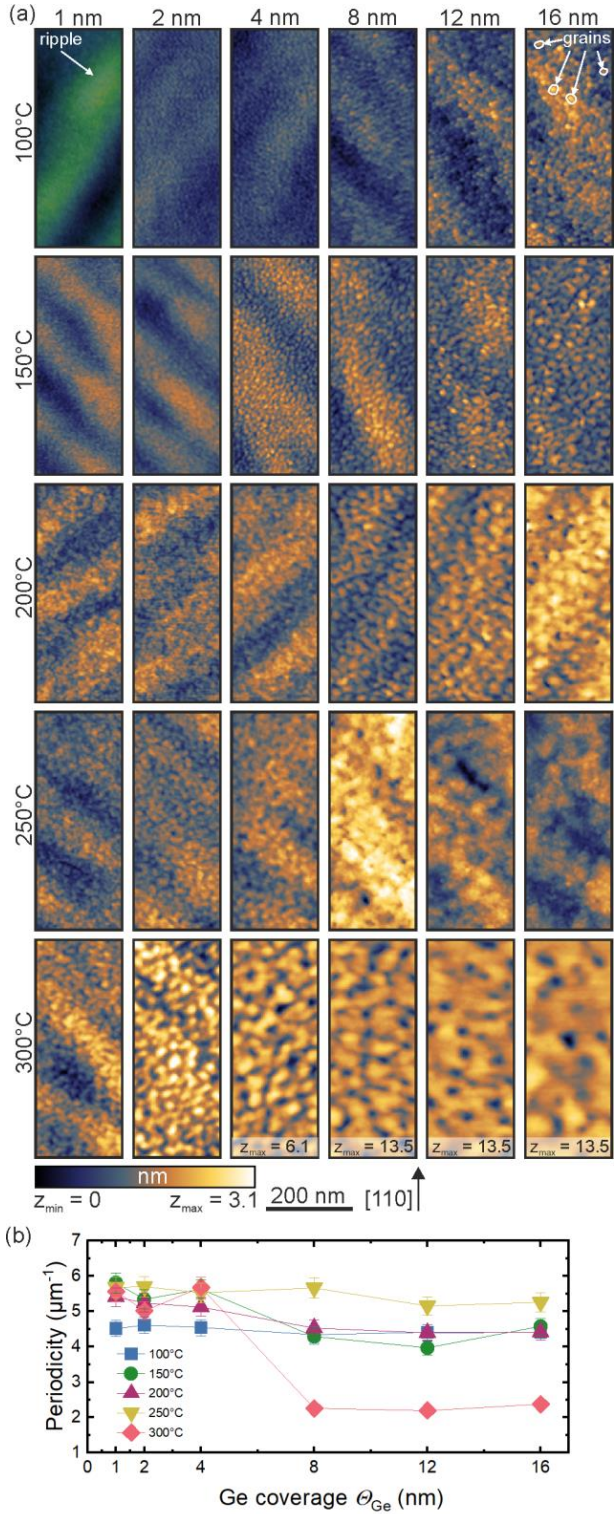

**Figure 2.** (a) 0.2×0.5 μm² close-ups of AFM images of all the strained samples, which have the same color coding except the samples 4 nm, 8 nm, 12 nm, and 16 nm grown at 300°C. (b) Periodicity of the ripples determined by 2D FFT of the 5×5 μm² micrographs. The error bars result from the analysis.

In this section, we present the experimental results of the $\Theta_{Ge}$ and $T_{Ge}$ series of Ge on Si(001), where the Ge rate was maintained at 0.005 nm/s. Following the deposition, the samples were inspected using AFM to investigate the morphological surface variations. Figure 2(a) provides close-ups (0.2×0.5 µm²) of the entire sample matrix, illustrating that even $\Theta_{Ge}$s of 1 nm exert a profound impact on the surface topography, in comparison to the vicinal surface of the high-$T$ Si buffer replicated from the Si wafer (see Figure S. 4(a)). Indeed, we observe a pronounced surface ripple feature, highlighted in green in the image for a $\Theta_{Ge} = 1$ nm at $T_{Ge} = 100$°C, across nearly all samples. The origin of this surface feature is the step-bunching of vicinal surfaces, although at the low $T_{Ge}$ employed here, they have a less regular appearance than previously demonstrated [44-47]. We analyzed the ripples by applying 2D FFTs to the 5×5 µm² micrographs and determined the peak positions by Gaussian fitting. The periodicity of the ripples as a function of $\Theta_{Ge}$ and $T_{Ge}$ (see Figure 2(b)) demonstrates no discernible trend for the majority of the samples, exhibiting a value of around 5 µm$^{-1}$. However, for the samples deposited at $T_{Ge} = 300$°C with $\Theta_{Ge} = 8$, 12, and 16 nm, we observe the disappearance of the ripples, as the periodicity falls to values below 2.5 µm$^{-1}$. Instead, higher aspect ratio "mound-like" features emerge, which we attribute to enhanced ad-atom energy enabling their redistribution on the surface, reminding of the Stranski-Krastanow dynamics. This is also confirmed by the root mean square (RMS) roughness values, which exceeded those of all the other samples and ranged from 0.8 nm to 1.5 nm. In addition, superimposed to the ripples, smaller grains are observed (see arrows in the figure of the sample with $\Theta_{Ge} = 16$ nm at a $T_{Ge} = 100$°C). For higher $T_{Ge}$ and $\Theta_{Ge}$, these grains increase in their lateral dimension. We note that for ultra-low growth temperatures (e.g., 100°C), a rather smooth surface with an RMS roughness ranging from 0.24 nm to 0.36 nm was observed, i.e., close to that of pristine Si(001) wafers.

Based on the obtained changes in the surface morphology indicating an influence of $T_{Ge}$ selected samples have been investigated via TEM. Figure 3(a) and (b) display DF images recorded near a $2\bar{2}0$ two beam condition (TBC) of the thin Ge layer. The perfect TBC is set at the position of the bright fringes, where excitation errors are nonexistent. The contrast is therefore a measure of the $2\bar{2}0$ plane orientation variations. The planes are perfectly aligned with respect to the incident beam at the bright fringe positions, while between the fringes the planes tilt away and back again to produce the given periodicity of the fringes. For the sample with a $\Theta_{Ge} = 4$ nm deposited at $T_{Ge} = 100$°C, we estimate about 10.6 fringes per 100 nm leading to an estimated 9.4 nm period; for the sample with an equal $\Theta_{Ge}$ but grown at $T_{Ge} = 250$°C around 6.5 fringes per 100 nm are present forming a larger periodicity of about 15.4 nm. Here is to mention, that usually DF images at the $2\bar{2}0$ TBC conditions are sensitive to misfit dislocation and their contrast should be visible However, no clear dislocation contrast could be observed, which is attributed to the reason that the misfit dislocations are probably only present in short segments and end in internal (point) defects. There may be also other misfit defects that are not visible under the imaging conditions selected for the plan view specimens. Perfect Ge regions without misfit dislocation can be found in our samples, as displayed in Figure 3(c) and (d) for a $\Theta_{Ge}$ = 4 nm grown at $T_{Ge}$ = 100°C and 250°C, respectively. However, these areas are separated by crystal sections containing extended defects with 60° perfect dislocations and twin/Σ9 defects [48]. These examples can be seen in Figure 3(e) and (f). This proves that misfit dislocations are clearly present and typically built up by two perfect 60° dislocations, as reported already for low $T_{Ge}$ epitaxy [49]. If the two 60° dislocations can combine by the given thermic budget to a sessile Lomer dislocation or remain two separated 60° dislocations cannot be unambiguously determined from these uncorrected phase contrast images. The other structure, the extrinsic twin/Σ9 defect, arises during the coalescence of two Ge growth nuclei [48].

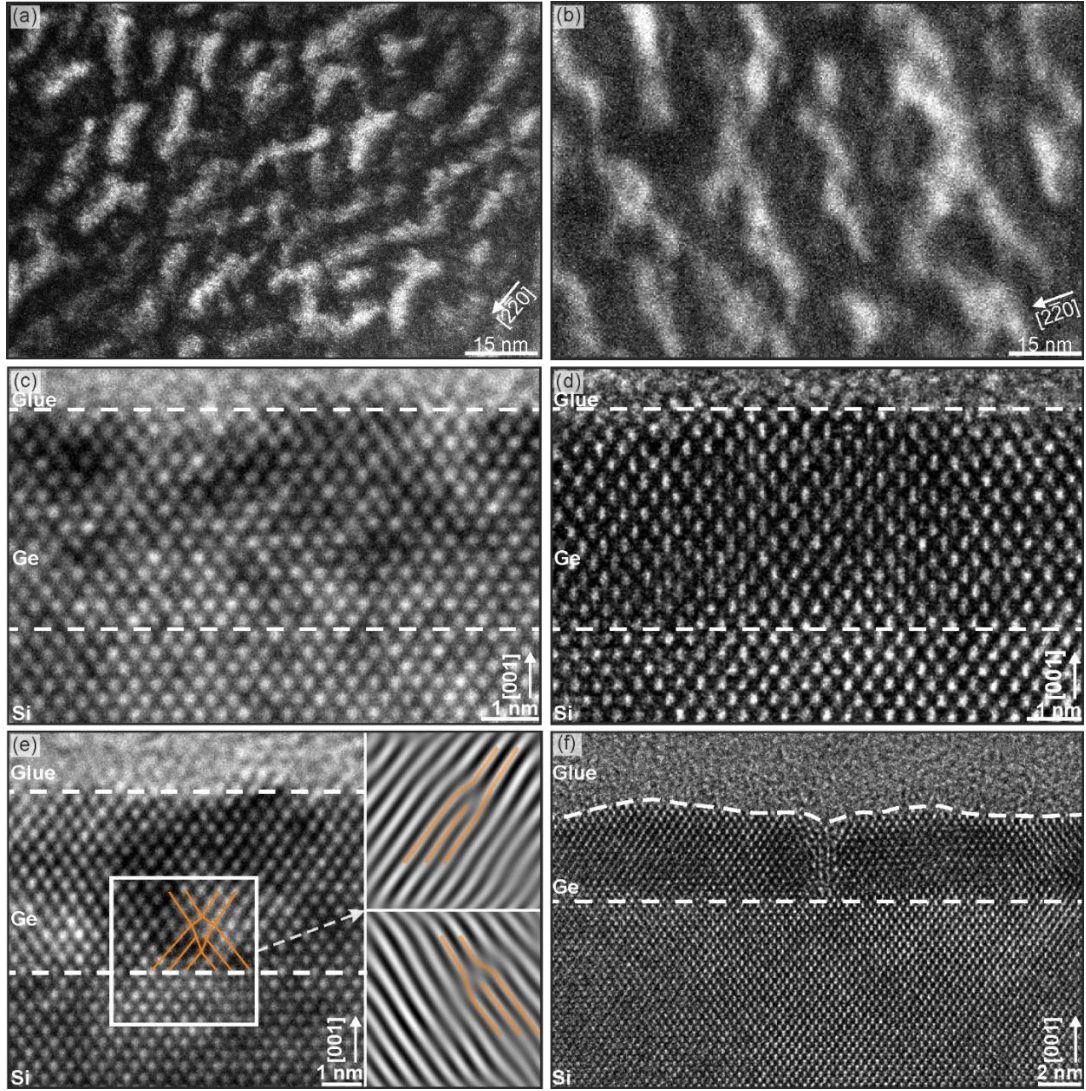

**Figure 3.** (a) Plan-view TEM from the 4nm/100°C sample recorded as DF at a $2\bar{2}0$ TBC sensitive to $2\bar{2}0$ plane variations. (b) Plan-view TEM of the 4nm/250°C sample recorded as DF at a $2\bar{2}0$ two-beam condition sensitive to $2\bar{2}0$ plane variations. (c, e) Cross-section HRTEM from the 4nm/100°C sample (d, f) Cross-section HRTEM 4nm/250°C sample.

The lattice strain of the epitaxial layers was systematically investigated by XRD measurements, allowing the tracking of the heteroepitaxial strain in dependence of $\Theta_{Ge}$ and $T_{Ge}$. Figure 4(a) presents a RSM around the 004 Bragg reflections of a sample with $\Theta_{Ge}$ = 4 nm, and $T_{Ge}$ = 100°C. We observe an intense peak from the Si substrate at an out-of-plane momentum transfer $Q_z$ = 7.365 nm$^{-1}$. Below, there is the signal from the Ge layer, which is elongated along $Q_z$ as the thin film comprises only ~32 atomic layers. Along the in-plane momentum transfer $Q_x$, which for a symmetric reflection corresponds to the rocking angle $\omega$, this Bragg peak is narrow, since the epilayer consists of pseudomorphic grains. Moreover, we observe fringes stemming from intensity modulation along $Q_z$ due to interference of x-rays reflected from the top and bottom surfaces of the Ge layer. A RSM for the $\Theta_{Ge} = 8$ nm layer deposited at $T_{Ge} = 100°C$ is shown in Figure 4(b). Interestingly, for this thicker layer, we observe a 'halo' of diffuse scattering along the sharp peak. This phenomenon is attributed to the onset of plastic relaxation, e.g. misfit dislocations limited to a certain amount of grains and the growth on top of twin/Σ9 defects. Both of these processes release the global biaxial strain in the Ge by inducing local fields of strain and rotation. This corresponds to a broadening of the spatial distribution of the lattice spacing, which leads to diffuse x-ray scattering. While the position of the 004 Bragg reflection is sensitive only to the out-of-plane strain, asymmetric Bragg reflections allow us to understand the 3D deformation of the Ge unit cell. In Figure 4(c) and (d), we show 224 RSMs for the two samples. Interestingly, for the 4 nm Ge layer in panel (c), we find two Bragg peaks stemming from the Ge layer. One is a sharp signal at the in-plane momentum transfer of the Si substrate at $Q_x = 5.205$ nm$^{-1}$, as the metastable Ge epilayer adapts its in-plane lattice parameter. However, there is also a secondary, diffuse peak shifted towards smaller $Q_x$ and larger $Q_z$. The simultaneous presence of the two peaks indicates that within this thin layer, some amount of Ge remains perfectly pseudomorphic, while other domains are distorted and partially relaxed, indicating structural defects. This observation is in agreement with what can be concluded from the TEM measurements carried out on the same sample (see Figure 3(a),(c),(e)), which revealed the presence of misfit dislocations and domain formation due to the coalescence of two grains. For the 8 nm-thick Ge layer (see Figure 4(d)), the Bragg peak from the pseudomorphic material is less intense, while the signal for the relaxed layer becomes even more diffuse and moves towards the position for a cubic lattice, indicated by the red line. From the linear combination of the 004 and 224 Bragg peaks, we calculate the lattice strains $\varepsilon_{xx} = \varepsilon_{yy}$ and $\varepsilon_{zz}$ for the partially relaxed layer peaks by the equations provided in [50], assuming the epilayer as pure Ge. In this way, we calculate the strains of the domains for a series of samples listed in Table 1, excluding the pseudomorphic regions.

**Table 1.** Strain state of the Ge domains determined by XRD.

| Sample | $\varepsilon_{xx}$ [%] | $\varepsilon_{zz}$ [%] | $v_{13}$ | $R_D$ [%] |
|---|---|---|---|---|
| 8 nm 100°C | -1.53 | 0.89 | 0.226 | 61.9 |
| 8 nm 150°C | -1.39 | 0.81 | 0.224 | 65.3 |
| 8 nm 200°C | -1.41 | 0.85 | 0.232 | 64.9 |
| 8 nm 250°C | -1.26 | 0.75 | 0.229 | 68.5 |
| 12 nm 150°C | -1.12 | 0.62 | 0.217 | 72.0 |
| 12 nm 200°C | -1.15 | 0.68 | 0.230 | 71.4 |
| 16 nm 100°C | -1.04 | 0.55 | 0.209 | 74.0 |
| 16 nm 150°C | -0.91 | 0.55 | 0.231 | 77.4 |

As expected, the strain decreases with increasing layer thickness and growth temperature as the metastable epilayer undergoes additional plastic relaxation when the thermal activation energy is available. We note that for the peaks of the pseudomorphic material, the in-plane strain is set by the lattice mismatch of Ge to the Si substrate, i.e. $\varepsilon_{xx} \cong 4.0$ %. Moreover, we determine the Poisson number by equation (2), finding that in all samples, it is smaller than the literature value of ~0.273. This is consistent with previous observations that the Poisson number in highly strained epitaxial thin films is smaller than in bulk materials [51], which may be attributed either to a change in the elastic parameters or the breakdown of linear elastic theory for non-infinitesimal strains [52]. When considering the degree of relaxation for each sample in Table 1, it is apparent that the average degree of relaxation $R_D$ of the domains is increasing with $\Theta_{Ge}$. Except for the layers at a $T_{Ge}$ = 150°C, the trend of $R_D$ with increased $T_{Ge}$ is comparable. In Figure 4(e)-(h), we show reciprocal space around the 224 Bragg diffraction from the Ge layers for the $\Theta_{Ge}$ = 4 nm sample series at different $T_{Ge}$ = 100°C – 250°C. We observe that with larger $T_{Ge}$, the intensity from the pseudomorphic peak decreases, while the diffuse signal stemming from the defective regions becomes more prominent. In Figure 4(i) and (j), rocking curves are plotted in dependence of the rocking angle $\omega$ across the 224 Ge peaks for the 4 nm and 8 nm sample series. Also, here, we observe that the ratio of the intensity of the pseudomorphic peak to that of the diffuse peak is decreasing with $T_{Ge}$ and $\Theta_{Ge}$, as more of the material undergoes relaxation. Moreover, for the 8 nm samples grown at $T_{Ge} > 100°C$, the formation of a secondary peak in the defective regions is apparent, corresponding to different degrees of relaxation within the same layer.

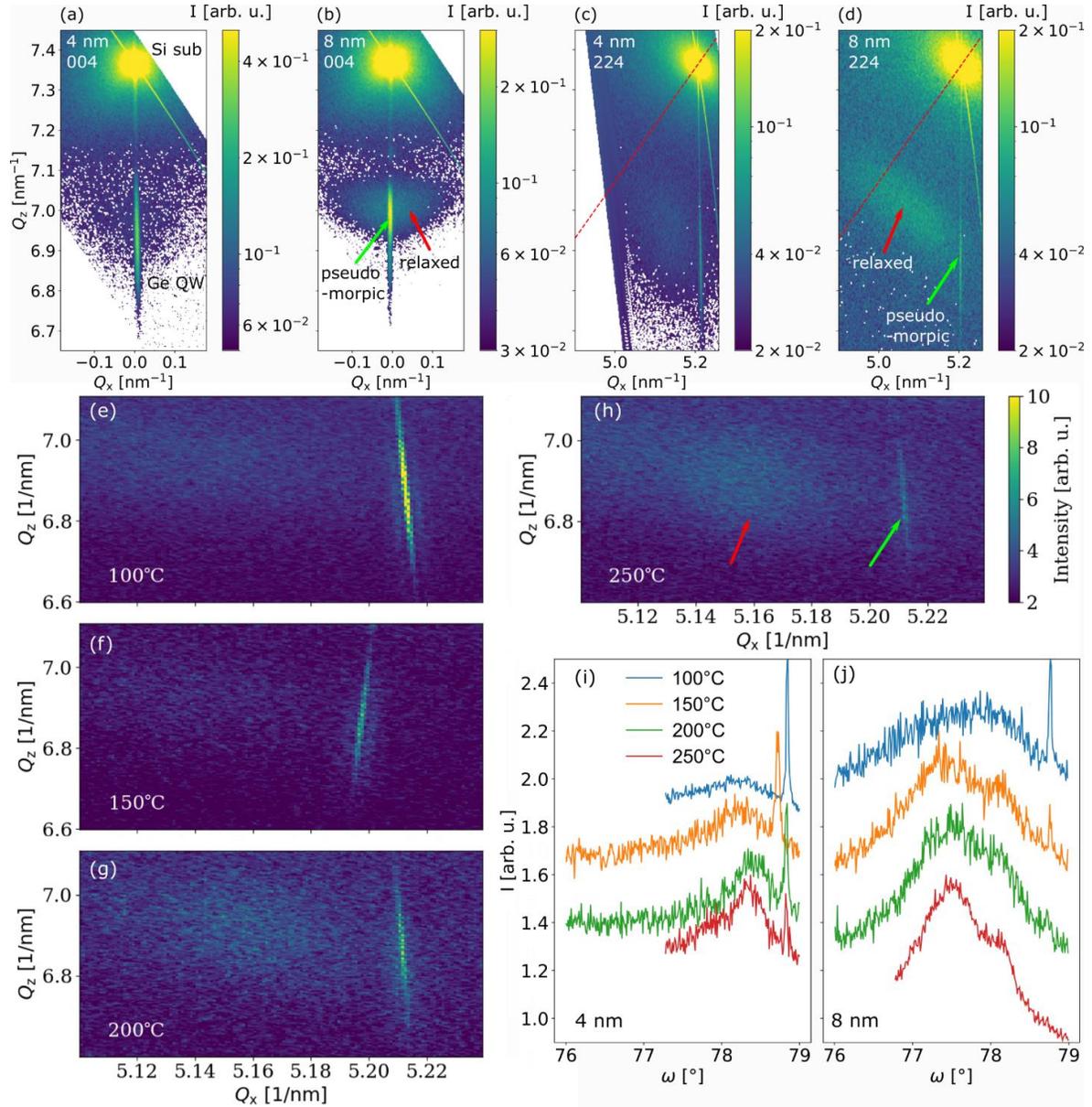

**Figure 4.** Characterization of ultra-thin epitaxial Ge/Si layers by XRD. (a) 004 RSM of 4 nm Ge layer grown at $T = 100°C$. (b) 004 RSM of 8 nm Ge layer at 100°C, the red and green arrows indicate the intensity diffracted from relaxed and pseudomorphic regions, respectively. (c) 224 RSM for 4 nm Ge at 100°C. The red line is the direction indicating a perfect cubic lattice. (d) 224 RSM for 8 nm Ge at 100 °C. (e-h) 224 RSMs around the Ge signal for 4 nm Ge layers grown at different temperatures. (i, j) 224 rocking curves across the Ge signals for 4 nm and 8 nm thick layers.

In order to study the spatial strain fluctuations, we investigated the 4 nm Ge layer grown at 250°C by SXDM [39], at the hard x-ray nanoprobe beamline ID01/ESRF [40], as sketched in Figure 5(a). Thus, we obtain a spatial map of $\varepsilon_{zz}$ with ~50 nm resolution, presented in Figure 5(b). The dominant features in the map take the form of semi-regular undulations reminiscent of the domains observed in the AFM images. Furthermore, the map of $w_{yz}$ obtained from the SXDM data is presented in Figure 5(c), showing a regular pattern of step-like undulations.

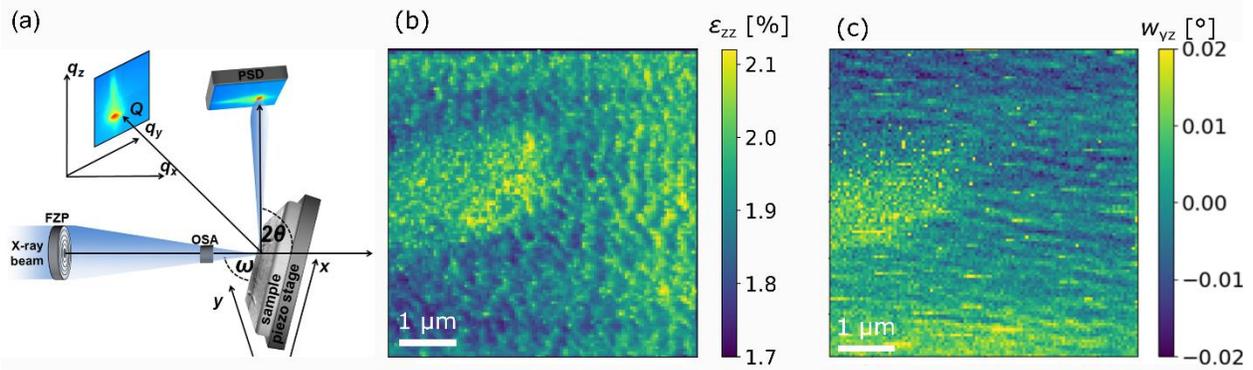

**Figure 5.** (a) Schematic setup for SXDM experiments at ID01/ESRF, reproduced with permission from [20]. (b) SXDM map of the $\varepsilon_{zz}$ strain component. (c) SXDM map of the $w_{yz}$ lattice rotation.

The TEM and SXDM results indicate the presence of strain relaxations on different length scales. To gain a comprehensive understanding of the surface morphology, Figure 6 presents a comparison of the measurements for the sample with a $\Theta_{Ge} = 4$ nm at $T_{Ge} = 250$°C. Figure 6(a) depicts an AFM height image (5×2.5 µm$^2$) and illustrates the larger surface structure. These ripples correspond well with the observed structure in the SXDM $\varepsilon_{zz}$ strain map (see Figure 6(d)). As previously discussed, the sample's surface also exhibits granular features, which can be observed in Figure 6(b). It shows an AFM image of a smaller size (1×0.5 µm$^2$). The inset of this AFM image and the plan-view TEM DF image in Figure 6(c) and (e), respectively, complete the picture of the surface. Although the size of the structures observed in the AFM image is slightly larger than the size of the TEM structures due to the lower resolution of the AFM, the results agree very well.

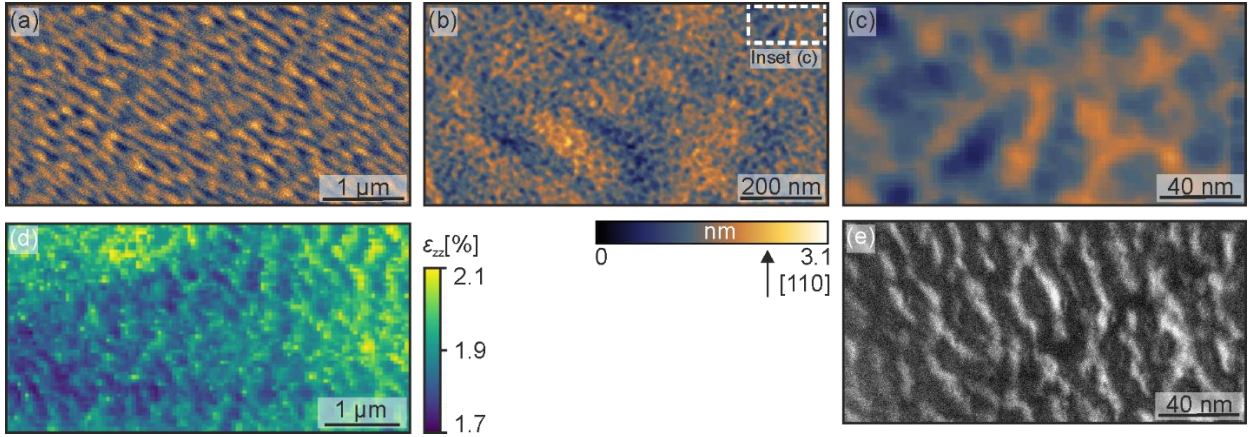

**Figure 6.** Comparison of the measurements for the sample with a $\Theta_{Ge}$ = 4 nm deposited at $T_{Ge}$ = 250°C. (a) 5×2.5 µm² AFM image showing the surface ripples. (b) 1×0.5 µm² AFM micrograph depicting ripples and superimposed grains. (c) Inset of (b) putting the focus on the grains. (d) SXDM map of the $\varepsilon_{zz}$ strain component. (e) Plan-view TEM DF image sensitive to $2\bar{2}0$ plane variations.

The reduction in the appearance of ripples at higher $T_{Ge}$ (=300°C) is consistent with previous studies that have demonstrated that step-bunching is kinetically driven and occurs in Si homoepitaxy without strain, too [53,54]. We have additionally demonstrated the formation of ripples in a 50 nm thick Si epilayer grown at 200°C on a high-$T$ Si buffer (see Figure S. 4(b)). Notably, the morphology of the Si ripples differs from that observed in the Ge samples, suggesting a strain-related contribution to the formation process. This observation aligns with the SXDM results, which indicate that the ripples influence the strain distribution within the sample.

Our findings of a granular surface corroborate the results of Storozhevyk et al., demonstrating that under a flux of Ge atoms at ULT, no islands form; instead, a surface composed of pseudomorphic Ge clusters emerged [55]. The kinetic limitations at ULT prevent the formation of larger islands, leading to the coalescence of Ge clusters into grown-in twin/Σ9 defects, which relaxes strain (see Figure 3(f)) [48]. A notable feature is that even with increased Ge deposition, the epilayers remained island-free and crystalline. Even at very low $T_{Ge}$, the grains formed were all pseudomorphic, resulting in a more intense XRD peak at the position of the Si substrate peak in the 224 RSM. Incoming Ge atoms diffused to the nearest relaxed position, typically the defect between two grains. At higher $T_{Ge}$, the adatoms have more energy, allowing for longer diffusion distances. Consequently, the grains broadened for both more $\Theta_{Ge}$ and higher $T_{Ge}$, with Ge atoms positioning within a range of 0.543 – 0.566 nm, corresponding to the lattice constants of Si and Ge, respectively. This broad distribution results in the diffuse signal observed in XRD. We note that for ultra-low temperature growth at excellent $p_{Ge}$s, the absence of epitaxial growth breakdown via the formation of amorphous layers represents a crucial finding. It clearly shows that epitaxial Ge layers can be grown with a thickness relevant for nanoelectronics applications. Indeed, 4 nm-thick Ge layers deposited using similar growth conditions, were recently implemented on thin silicon-on-insulator substrates for nanoelectronic device applications, such as reconfigurable transistors [25-27]. These devices clearly outperformed reference devices based on pure SOI, Ge-on-insulator substrates [26] and devices based on harvested Ge VLS nanowires. Interestingly, in these Si/Ge/Si on -insulator nanosheets, no defects such as the here observed grown-in twin/Σ9 defects were found, even if the Ge layers had been grown on the Si device layer at similar $\Theta_{Ge}$ and $T_{Ge}$. We note that the whole Si/Ge/Si nanosheet structure on the insulator was very thin in [24-27], i.e., < 35 nm thick. Thus, effects like strain partitioning (compliance) between Si and Ge layers [56] and the role of tensile strain induced by the Si/SiO$_2$ interface after thermal oxidation [57] could influence the relaxation dynamics of the Ge layers. Consequently, the difference in the growth of Ge layers on bulk Si and thin-SOI has to be further investigated.

## 2.4. Conclusion

We investigated the growth characteristics of Ge layers on Ge(001) and Ge layers on Si(001) using MBE at ultra-low-growth temperatures by varying the layer thickness well beyond the limits for elastic relaxation at conventional high growth temperatures. For the homoepitaxy, VEPALS investigations demonstrate that even at $T_{Ge}$ = 100°C, highly crystalline growth is possible if the background pressure during the growth is kept low. For the strained heteroepitaxy, AFM revealed surface ripples with superimposed grains for $T_{Ge}$ < 300°C or small $\Theta_{Ge}$s at $T_{Ge}$ = 300°C. Plan-view DF TEM and HRTEM showed that these grains coalesce into twin/Σ9 defects, relieving misfit stress with misfit dislocations. Despite these defects, the Ge layers within grains exhibited high crystalline quality, pseudomorphic characteristics, and expected strain. XRD measurements confirmed both distorted and crystalline growth through two distinct peaks. The strain state and the $R_D$ of the relaxed regions exhibited an increase with $\Theta_{Ge}$ and $T_{Ge}$. The kinetic limitations of the ULT growth were identified as the cause of the ripples, with strain influencing their formation. This was demonstrated by nanobeam x-ray diffraction measurements, which indicated strain variations in regions with and without ripples. The incorporation of the aforementioned defects formed upon epitaxy of Ge on Si is rather a result of the high strain in the layers than of the ultra-low growth temperature per se. Thus, strain management through, e.g., the use of strained-SOI substrates [24] or selective epitaxy on ridges [58,59] or nanotips [56] combined with ULT growth in deep UHV can be the route for implementing high-quality, supersaturated Ge layers on Si for novel device applications. These findings of Ge growth at ultra-low-growth temperatures contribute to the feasibility of scalable top-down fabrication techniques for Ge-based nanoelectronic devices [26,27]. This development opens up potential opportunities for further technological advancements in next-generation semiconductor technologies.

## ASSOCIATED CONTENT

## AUTHOR INFORMATION

**Corresponding Author**

*Christoph Wilflingseder - Institute of Semiconductor and Solid State Physics, Johannes Kepler University, Altenberger Straße 69, 4040, Linz, Austria – christoph.wilflingseder@jku.at

**Author Contributions**

The manuscript was written through contributions of all authors. All authors have given approval to the final version of the manuscript.

**Notes**

**Acknowledgment**

This research was funded in whole or in part by the Austrian Science Fund (FWF) [10.55776/Y1238]. For open access purposes, the author has applied a CC BY public copyright license to any author-accepted manuscript version arising from this submission. We acknowledge the European Synchrotron Radiation Facility (ESRF) for provision of synchrotron radiation facilities and we would like to thank the staff for assistance in using beamline ID01. The financial support by the Austrian Federal Ministry of Labour and Economy, the National Foundation for Research, Technology and Development, and the Christian Doppler Research Association is gratefully acknowledged. Parts of this research were carried out at ELBE at the Helmholtz-Zentrum Dresden - Rossendorf e. V., a member of the Helmholtz Association. We would like to thank the facility staff for their assistance.

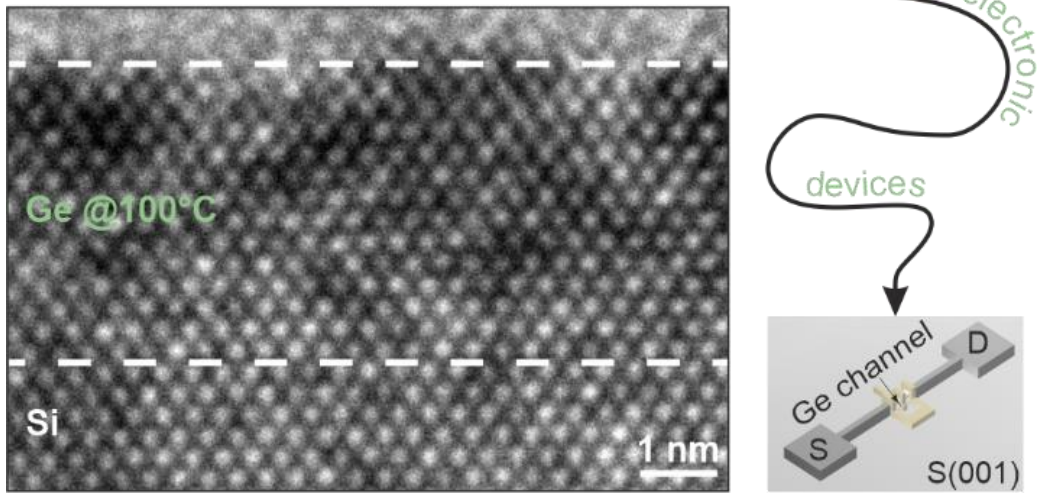

For Table of Contents only.

**Supporting Information**

Ge epitaxy at ultra-low growth temperatures enabled by a pristine growth environment

*Christoph Wilflingseder, Johannes Aberl, Enrique Prado Navarette, Günter Hesser, Heiko Groiss, Maciej O. Liedke, Maik Butterling, Andreas Wagner, Eric Hirschmann, Cedric Corley-Wiciak, Marvin H. Zoellner, Giovanni Capellini, Thomas Fromherz, Moritz Brehm*

Typically, high $T_{Ge}$ > 400°C is employed to ensure the growth of Ge on Si with good crystallinity. For ultra-low $T_{Ge}$s (ULT ≡ <300°C for Ge on Si), the ad-atom surface diffusion on Si and the Ge WL at low $T_{Ge}$ is reduced, and impurities originating from the chamber background and the sources can be incorporated during growth [34]. Therefore, Figure S. 1 emphasizes the importance of stringent vacuum conditions. In deep UHV, the $T_{Ge}$ exhibits minimal influence on the crystallinity (see Figure S. 1(a))). In contrast, higher growth pressures lead to an enhanced impingement rate of residual gases [60] (see Figure S. 1(c)). These impurities cannot be desorbed [34] and the results are amorphization or defective growth, which is depicted in the scheme in Figure S 1(b).

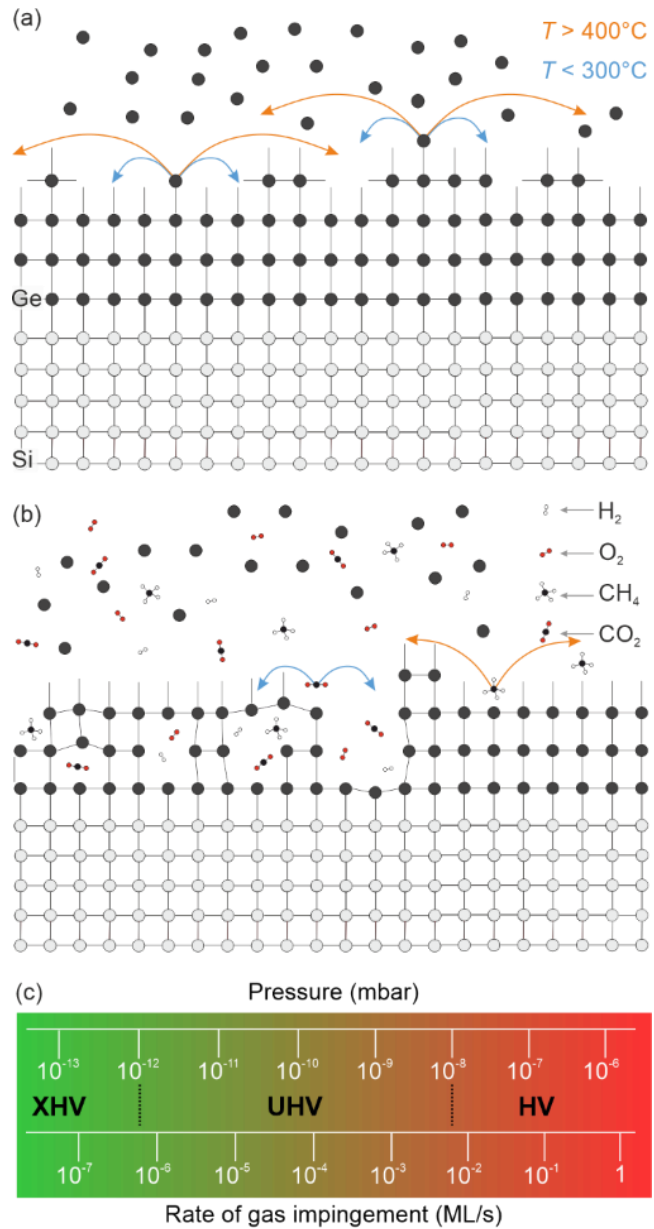

**Figure S. 1.** Scheme of the growth of Ge on Si at low $T_{Ge}$. The arrows indicate the surface diffusion at higher $T_{Ge}$ (orange) and lower $T_{Ge}$ (blue). (a) High epitaxial quality growth due to deep UHV growth conditions, i.e., mobility-limited growth. (b) Epitaxial growth at higher pressures, i.e., impurity-limited growth. Depending on the $T_{Ge}$ and the possibility of impurity desorption or the lack thereof, crystalline growth or amorphization occurs. (c) relationship between growth pressure and rate of gas impingement (ML/s), see Ref. [60].

As the $p_{Ge}$ is a critical parameter for ULT deposition, the growth chamber underwent a series of extensive preparations, including chamber conditioning, degassing, pumping, and the implementation of several gettering strategies. As a consequence, UHV pressure was achieved and the log files depicted in Figure S. 2(a) and (b) present the lowest and the highest $p_{Ge}$ observed during Ge deposition. Additionally, the shutter status, TGs, and growth rates are also included. The reference values for $p_{Ge}$ are taken when the deposition of half the total Ge layer has been completed and are approximately $\sim 9 \cdot 10^{-11}$ mbar and $\sim 2 \cdot 10^{-10}$ mbar.

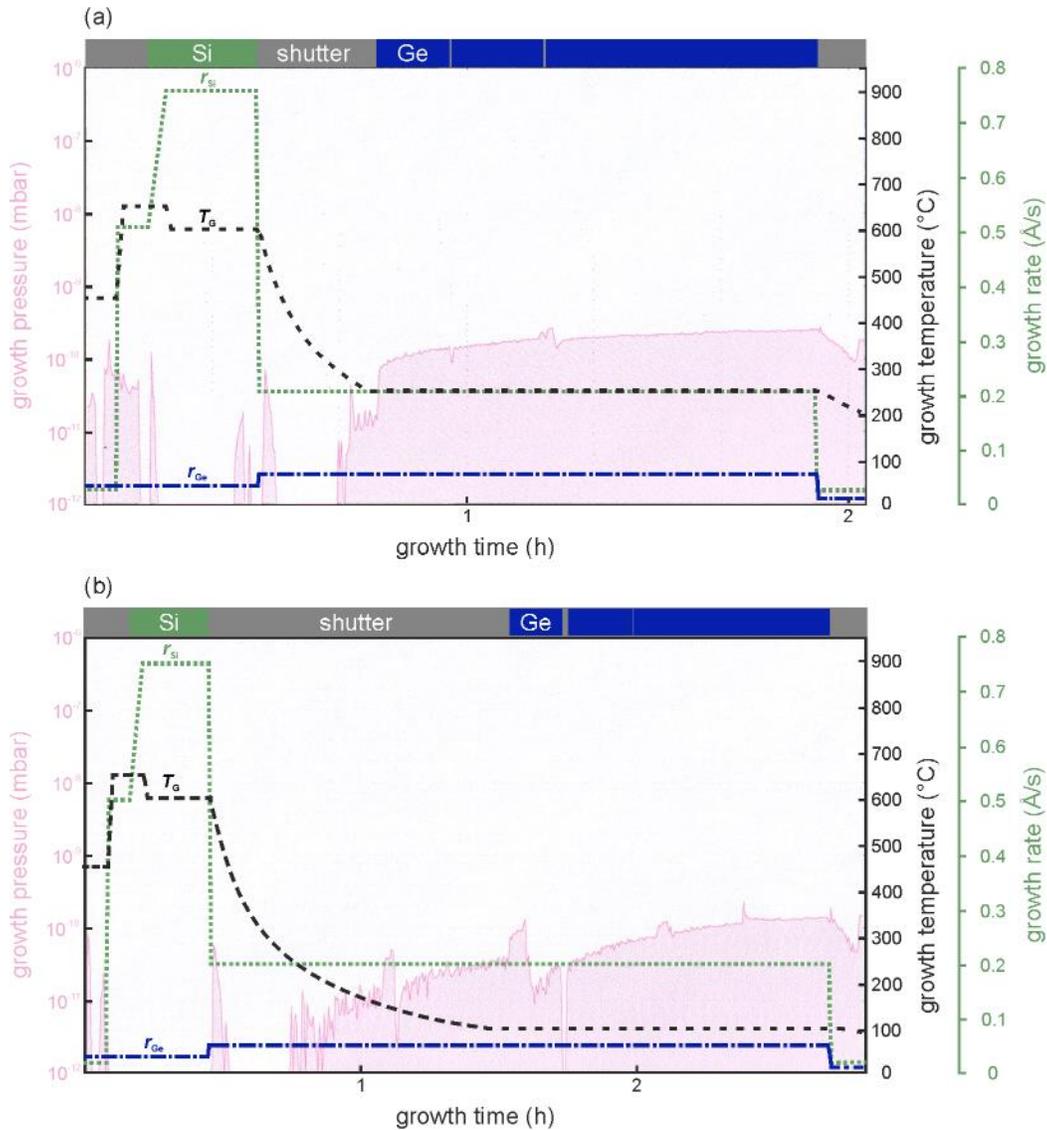

**Figure S. 2.** Growth log files. (a) 8 nm, 12 nm and 16 nm at $T_{Ge}$=250°C. (b) 8 nm, 12 nm und 16 nm Ge at $T_{Ge}$=100°C.

Figure S. 3 provides AFM micrographs with dimensions of 5×2.5 μm², which were recorded to investigate larger features. The ripples occur very regularly and exhibit a preferential orientation, occurring either along the [100] or [010] crystal direction. This finding aligns with previous research, which also demonstrated that the ripples are oriented along these directions [61,62]. Qualitatively, the ripples diminish for a $T_{Ge}$ = 300°C and $\Theta_{Ge}s \geq 8$ nm. Given the numerous ripples present in these micrographs, they were used for the analysis of the periodicity.

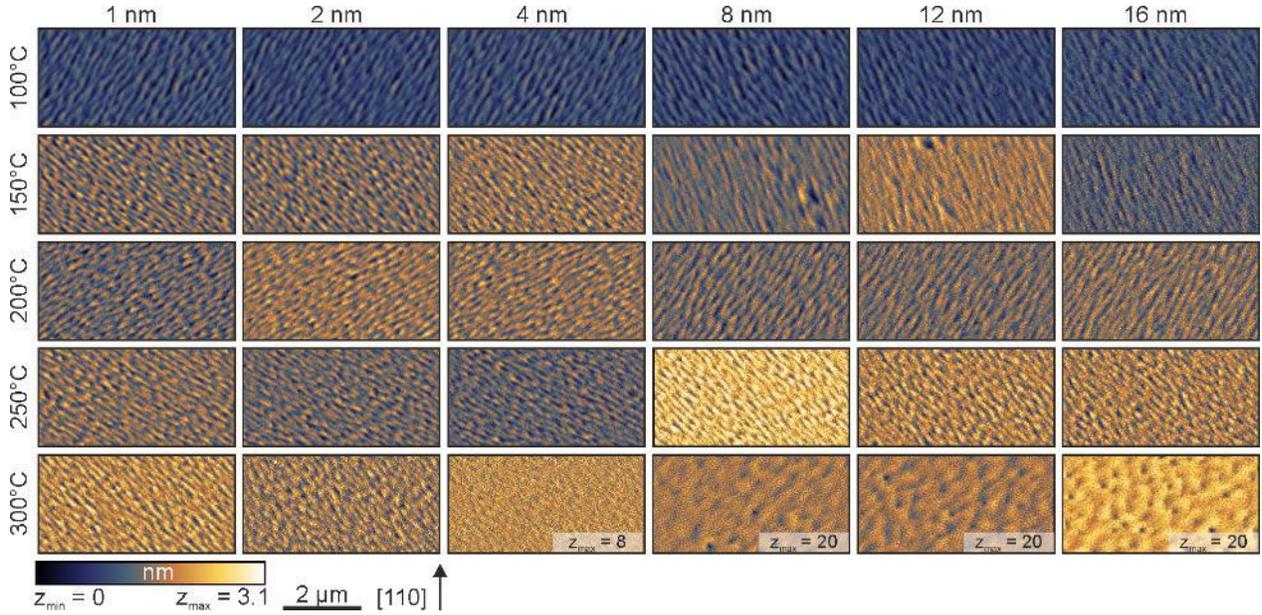

**Figure S. 3.** 5×2.5 μm² AFM images of the grown matrix.

In addition to the Ge/Ge(001) and Ge/Si(001) series, two complementary samples were grown. One sample was prepared by regrowing only the high-$T$, 75.5 nm thick Si buffer layer used in the heteroepitaxy study, and the resulting surface topography is shown in Figure S. 4(a). The vicinal surface of the Si(001) substrate was replicated. In the second sample, another Si homoepitaxy was conducted, with 50 nm of Si deposited at 200°C on a 50 nm thick Si buffer with a growth temperature of 650°C. As observed previously, ripples appeared in a regular pattern along the <100> direction, exhibiting a periodicity of approximately 2.4 μm$^{-1}$ (see Figure S. 4(b)).

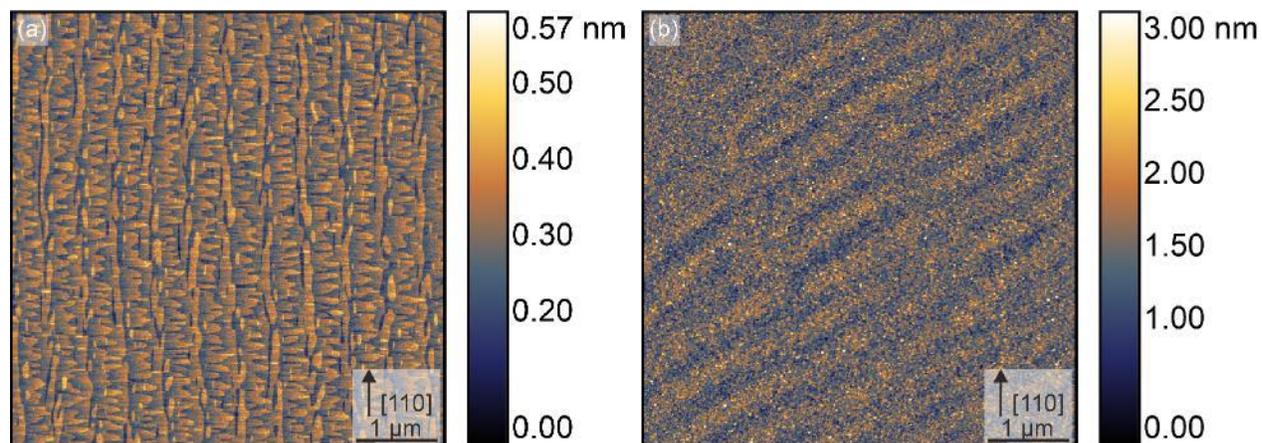

**Figure S. 4.** Complementary AFM images. (a) 5×5 μm² micrograph of the high temperature Si buffer. (b) 5×5 μm² micrograph of 50 nm Si homoepitaxy grown at 200°C.